\documentclass[12pt]{article}

\usepackage{epsfig}

\begin{document}

\renewcommand{\thefootnote}{\fnsymbol{footnote}}
\setcounter{page}{0}

\begin{titlepage}

\begin{flushright}
CERN-TH/98-277 \\
DESY 98-111 \\
RAL-TR-1998-062 \\
hep-ph/9808433 \\
August 1998
\end{flushright}

\vspace*{1.0cm}

\begin{center}
{\large \bf HIGGS RADIATION OFF TOP QUARKS} \\[0.5cm]
{\large \bf IN \boldmath{$e^+e^-$} COLLISIONS}\\
\end{center}

\vskip 1.cm
\begin{center}
{\sc S. Dittmaier$^1$, M. Kr\"amer$^2$, Y. Liao$^{3,4}$,} \\[0.3cm]
{\sc M. Spira$^5$ and P.M. Zerwas$^4$} 

\vskip 0.8cm

\begin{small} 
$^1$ Theory Division, CERN, CH-1211 Geneva 23, Switzerland          
\vskip 0.2cm
$^2$ CLRC Rutherford Appleton Laboratory, Chilton, Didcot, OX11 0QX, England
\vskip 0.2cm
$^3$ Department of Modern Applied Physics, Tsinghua University, 
Beijing 100084, PR China\footnote{Permanent address}
\vskip 0.2cm
$^4$  DESY, Deutsches Elektronen-Synchrotron, D-22603 Hamburg, Germany 
\vskip 0.2cm
$^5$ II.\ Institut f\"ur Theoretische Physik\footnote{Supported by
  Bundesministerium f\"ur Bildung und Forschung (BMBF), Bonn, Germany,
  under Contract 05~7~HH~92P~(5), and by EU Program {\it Human Capital
    and Mobility} through Network {\it Physics at High Energy
    Colliders} under Contract CHRX-CT93-0357 (DG12 COMA).},
Universit\"at Hamburg, D-22761 Hamburg, Germany
\end{small}
\end{center}

\vskip 2cm

\begin{abstract}
\noindent
The strength of the Yukawa coupling of the leptons and quarks with the
Higgs boson is uniquely determined by the Higgs mechanism for
generating masses in the Standard Model. The top-Higgs coupling can be
measured directly in the process $e^+e^- \rightarrow t \bar{t}H$.
This process therefore provides a fundamental test of the Higgs
mechanism. Extending earlier analyses carried out in the Born
approximation, we have determined the cross section for this process
including QCD corrections, which turn out to be important.
\end{abstract}

\end{titlepage}

\renewcommand{\thefootnote}{\arabic{footnote}}

\setcounter{footnote}{0}

\clearpage

\begin{figure}[p]
 \epsfysize=24cm
 \epsfxsize=18cm
 \centerline{\epsffile{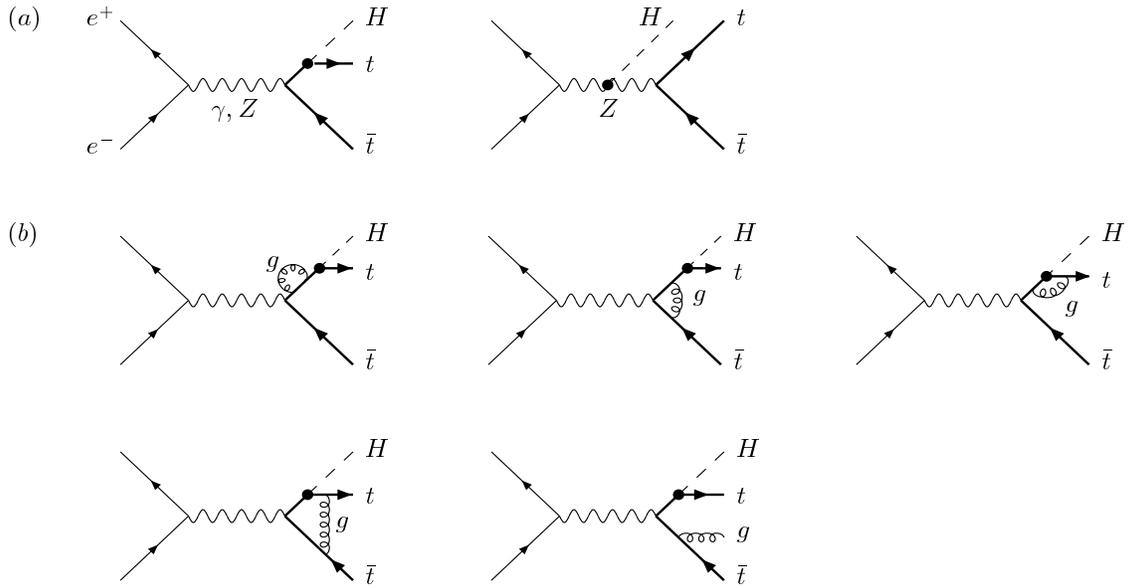}}
\vspace*{-11.5cm}
\caption[ ]{\it 
  Generic diagrams contributing to the process $e^+e^- \to t\bar t H
  (g)$ at (a) lowest and (b) next-to-leading order. The NLO set must
  be supplemented by the well-known QCD corrections to $Z^*\to
  t\bar{t}$, which are not depicted explicitly. }
\label{fig:diagrams}
\end{figure}

\begin{figure}[p]
 \vspace*{-2.5cm} 
 \epsfysize=12cm 
 \epsfxsize=12cm
 \centerline{\epsffile{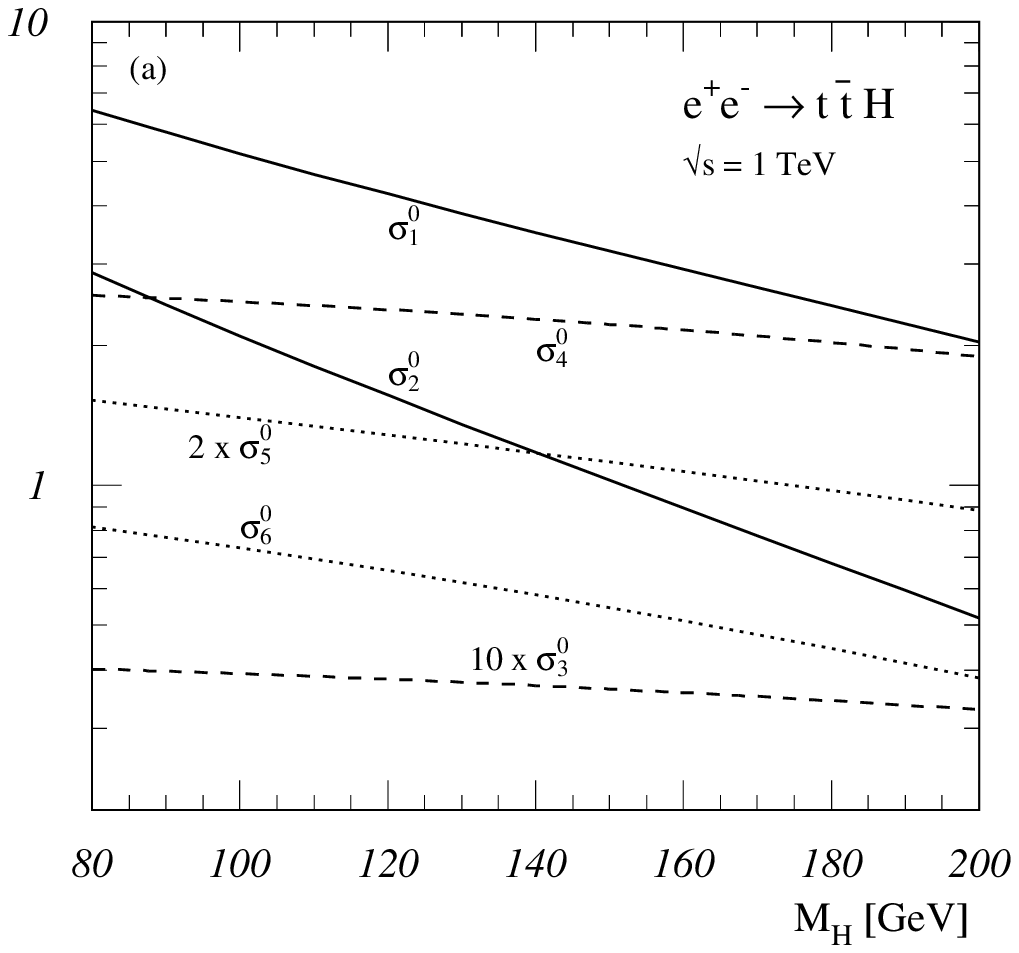}} 
 \vspace*{-2.5cm} 
 \epsfysize=12cm 
 \epsfxsize=12cm
 \centerline{\epsffile{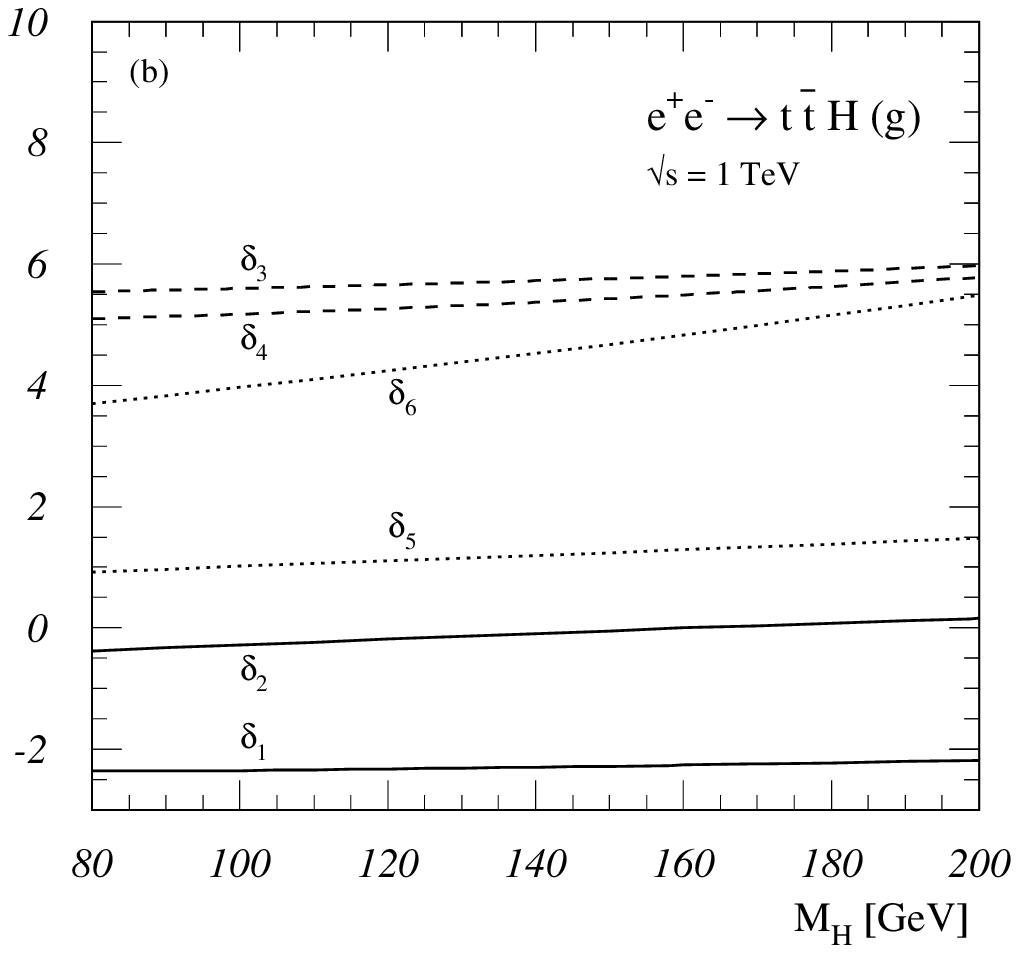}}
\vspace*{-0cm}
\caption[ ]{\it 
  (a) The vector and axial-vector contributions $\sigma_i^0
  (i=1,\ldots,6)$ of Higgs radiation off top quarks, off the $Z$
  boson, and the interference term in Born approximation; and (b) the
  QCD radiative corrections of these terms to order $\alpha_s$.}
\label{fig:coeffs}
\end{figure}

\begin{figure}[p]
 \epsfysize=12cm
 \epsfxsize=12cm
 \centerline{\epsffile{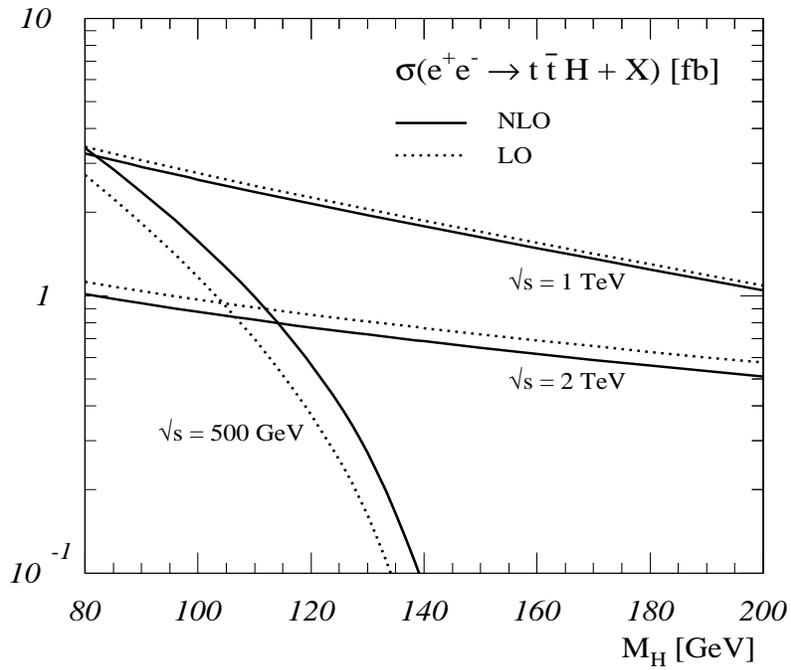}}
\vspace*{-0cm}
\caption[ ]{\it 
  The total cross section for the process $e^+e^-\to t\bar{t} H + X$,
  including QCD radiative corrections, for three collider energies as a
  function of the Higgs mass (full curves). The Born approximation
  (dotted curves) is shown for comparison.}
\label{fig:cs}
\end{figure}

\begin{figure}[p]
 \epsfysize=12cm
 \epsfxsize=12cm
 \centerline{\epsffile{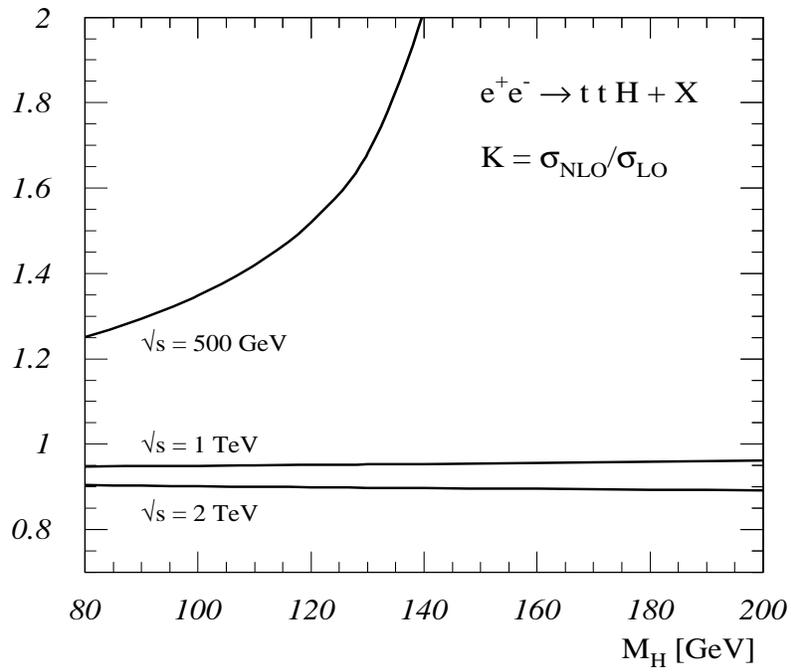}}
\vspace*{-0cm}
\caption[ ]{\it 
  The $K$ factor for the process $e^+e^-\to t\bar{t} H + X$ for three
  collider energies as a function of the Higgs mass.}
\label{fig:K}
\end{figure}

\noindent
1. The masses of the fundamental particles are generated in the Higgs
mechanism \cite{hi64} by interactions with the non-zero Higgs field in
the ground state. Within the Standard Model, the Yukawa couplings of
the leptons and quarks with the Higgs particle are therefore uniquely
determined by the particle masses:
\begin{equation}
g_{ffH} = m_f/v
\label{eq:gffh}
\end{equation}
with $v = (\sqrt{2} G_F)^{-1/2} \approx 246$~GeV. The measurement of
these couplings provides a fundamental experimental test of the Higgs
mechanism {\it sui generis}.
\\[1em]
\indent 
Several methods have been proposed in the literature to test Yukawa
couplings of the Higgs boson \cite{gu90}. In the present note we focus
on the top--Higgs Yukawa coupling. Since the mass of the top quark is
maximal within the fermion multiplets of the Standard Model, the $ttH$
coupling ranks among the most interesting predictions of the Higgs
mechanism. This coupling can be indirectly tested by measuring the
$H\gamma \gamma$ and $Hgg$ couplings, which are mediated by virtual
top-quark loops \cite{el76,el76a}.  If the Higgs boson is very heavy,
the decay mode $H \rightarrow t\bar{t}$ can be exploited \cite{ha91}.
On the other hand, if the Higgs boson is light, $M_H \sim
100$--200~GeV, the radiation of Higgs bosons off top quarks lends
itself as a basic mechanism for measuring the $ttH$ coupling. The
prospects of operating $e^+e^-$ linear colliders at high
luminosities\footnote{In TESLA designs, integrated luminosities of
$\int {\cal L}=0.3$~ab$^{-1}$ and $0.5$~ab$^{-1}$ per year are planned
at c.m.\ energies $\sqrt{s}=0.5$~TeV and $0.8$~TeV, respectively.}
\cite{ac98} render the radiation process \cite{ga78,dj92}
\begin{equation}
e^+ e^- \to t\bar{t} H
\end{equation}
a suitable instrument to carry out this measurement.
\\[1em]
\indent While for small top masses the top-quark pairs would have been
generated at lower energies by photon exchange \cite{ga78} and the
Higgs bosons would have been radiated exclusively off the top quarks,
the $Z$ exchange \cite{dj92} at high energies gives rise to an
additional small contribution from Higgs-strahlung. QCD corrections to
this process have recently been discussed in Ref.~\cite{da97}, where
the Higgs mass has been assumed small with respect to the top-quark
mass and the total energy.  Rescattering corrections, which are
potentially important at non-asymptotic energies, are not properly
taken account of by the fragmentation method used in
Ref.~\cite{da97}. In the present analysis we have performed a complete
calculation of the QCD radiative corrections to order $\alpha_s$ that
is applicable to all kinematical configurations.  For high energies,
in agreement with the earlier estimates \cite{da97}, the corrections
are negative. At modest energies, however, the corrections are
positive and large, and they increase the cross section significantly.
\\[2em]
2. The generic set of lowest-order diagrams for the radiation of Higgs
bosons off top quarks and Higgs-strahlung in $e^+e^-$ annihilation is
shown in Fig.~\ref{fig:diagrams}a. The QCD corrections can be
classified as virtual top-quark self-energy corrections, $tt\gamma$,
$ttZ$ and $ttH$ vertex corrections, box rescattering corrections and
real-gluon radiation, see Fig.~\ref{fig:diagrams}b. The total cross
section can be split into three parts: the cross section corresponding
to Higgs radiation off the top quarks ($\sigma_1$, $\sigma_2$),
Higgs-strahlung off the $Z$-boson line ($\sigma_3$, $\sigma_4$), and
the interference between these mechanisms ($\sigma_5$, $\sigma_6$).
Each of the contributions has been divided into two parts, generated
by the vector couplings of $\gamma$ and $Z$ to the top pair, and by
the axial-vector coupling of the $Z$.  These contributions are
different, owing to chirality-breaking mass terms and scalar vertices.
In a notation parallel to Ref.~\cite{dj92}, but slightly modified, the
total cross section may be written in the form
\begin{eqnarray}
\lefteqn{\sigma(e^+e^- \to t\bar t H (g) ) = } & 
\nonumber \\[.5em]
 & & N_{C} \frac{\sigma_0}{4\pi} \left\{ 
\frac{(\hat v_e^2 + \hat a_e^2)}{(1-M_Z^2/s)^2}
 \frac{g^2_{ttH}}{4\pi} 
\left( \hat v_t^2\sigma_1 +\hat a_t^2\sigma_2 \right)
+\left( Q_e^2Q_t^2 + \frac{2Q_eQ_t\hat v_e\hat v_t}{1-M_Z^2/s}
\right) \frac{g^2_{ttH}}{4\pi} \sigma_1 \right. 
\nonumber \\
& & \hspace*{1.5cm} {} + \frac{(\hat v_e^2 + \hat a_e^2)}{(1-M_Z^2/s)^2}
\frac{g^2_{ZZH}}{4\pi} \left( \hat v_t^2 \sigma_3 + \hat a_t^2 \sigma_4
\right) 
\nonumber \\
& & \hspace*{1.5cm} {}+ \left. \frac{(\hat v_e^2 + \hat a_e^2)}{(1-M_Z^2/s)^2}
\frac{g_{ttH}g_{ZZH}}{4\pi} \left( \hat v_t^2 \sigma_5 + \hat a_t^2 \sigma_6
\right)  + \frac{Q_eQ_t\hat v_e\hat v_t}{1-M_Z^2/s}
\frac{g_{ttH}g_{ZZH}}{4\pi} \sigma_5 \right\}.
\end{eqnarray}
The $ttH$ Yukawa coupling $g_{ttH}$, see Eq.~(\ref{eq:gffh}), and the
$ZZH$ coupling $g_{ZZH} = M_Z/v$ are set off explicitly.\footnote{In
the analysis below, $v$ is calculated from the electroweak input data
via $v=M_Z c_W s_W / \sqrt{\pi\alpha}$.}  The electric charges are
defined as $Q_e = -1$ and $Q_t = + 2/3$; $\hat{v}_{e,t}$ and
$\hat{a}_{e,t}$ are the vector and axial-vector charges of electron
and top quark, normalized as $\hat{v} = (2 I_{3L} - 4Q s^2_W) / (4c_W
s_W)$ and $\hat{a} = 2 I_{3L} / (4c_Ws_W)$, with $I_{3L} = \pm 1/2$
being the weak isospin of the left-handed fermions [as usual, $s^2_W =
1 -c^2_W = \sin^2 \theta_W = 0.23$]; $\sigma_0$ denotes the standard
electromagnetic $\mu$-pair cross section, $\sigma_0 = 4 \pi
\alpha^2/3s$, and $N_{C} = 3$ is the colour factor of the top quarks.
The value of the electromagnetic coupling is taken at $\alpha =
1/128$. The coefficients $\sigma_i (i = 1,\ldots, 6)$ can be
decomposed into Born contributions $\sigma^0_i$ and QCD corrections
$\delta_i$:
\begin{equation}\label{eq:sigi}
\sigma_i = \sigma_i^0 \left[ 1+ \frac{\alpha_s}{\pi} \delta_i \right],
\hspace{1cm} i=1,\ldots,6,
\end{equation}
where we choose $\mu^2_R = s$ for the renormalization
scale\footnote{The scale is fixed only to logarithmic
accuracy. Different choices of scales can readily be implemented in
the definition of the QCD corrections in Eq.~(\ref{eq:sigi}).}  of the
QCD coupling $\alpha_s$.  The total c.m.\ energy is denoted by
$\sqrt{s}$, and $M_Z=91.187$~GeV and $M_t = 174$~GeV \cite{pa98} are
the masses of the $Z$ boson and the top quark, respectively.
\\[1em]
\indent 
The QCD radiative corrections have been calculated in the standard
way. The Feynman diagrams and the amplitudes for the virtual
corrections have been generated with {\sl Feyn\-Arts} \cite{ku91}.
They have been evaluated by applying the standard techniques for
one-loop calculations, as described in
Ref.~\cite{th79,de93}. Ultraviolet divergences have been consistently
regularized in $D=4-2\epsilon$ dimensions, with $\gamma_5$ treated
naively.  Note that the renormalization of the $ttH$ vertex is
connected to the renormalization of the top-quark mass, which is
defined on shell (see Refs.~\cite{el76a,de93,dj96} for details).  The
algebraic part of the virtual corrections has also been checked by
using {\sl Feyn\-Calc} \cite{me91}.  The infrared divergences
encountered in the virtual corrections and in the cross section for
real-gluon emission, have been regularized in two ways, by using
dimensional regularization and by introducing an infinitesimal gluon
mass; both ways led to identical results after adding the
contributions from virtual gluon exchange and real-gluon emission. A
second, completely independent calculation of the QCD corrections to
the total cross section was based on the evaluation of all relevant
cut diagrams of the photon and $Z$-boson self-energies in two-loop
order, generalizing the method applied to $t\bar{t}(g)$ intermediate
states in Ref.~\cite{dg94}. The results of the two approaches are in
numerical agreement. Finally, we have performed a mutual comparison of
partial results with the authors of Ref.~\cite{da98}; agreement was
found between the calculations.
\\[1em]
3. The integrated Born coefficients $\sigma^0_i$ and the QCD
corrections $\delta_i$ are too complex to be recorded
analytically.\footnote{The Dalitz densities are given at the Born
level analytically in Ref.~\cite{dj92}. Note that the Higgs couplings
to fermions $g_{ffH} $ are defined with a minus sign in
Ref.~\cite{dj92}.} [The corresponding Fortran program is available
from DESY \cite{forprg}.] In this letter, we restrict ourselves to the
discussion of the basic numerical results for the total cross
section. To characterize the relative size of the contributions
$\sigma^0_i$, the Born coefficients are shown in
Fig.~\ref{fig:coeffs}a as functions of the Higgs mass for a total
energy $\sqrt{s} = 1$~TeV. The corresponding QCD corrections
$\delta_i$ are shown in Fig.~\ref{fig:coeffs}b; in the subsequent
evaluation of the cross section, the QCD coupling is evaluated at the
two-loop level, normalized to $\alpha_s (M^2_Z) = 0.119$ \cite{ca98}.
\\[1em]
\indent 
The size of the QCD corrections is large at modest energies. This is a
consequence of the rescattering diagrams, which are generated by the
gluon exchange between the final-state top quarks near the $t\bar t$
threshold. This Coulomb singularity leads to corrections of the form
$\alpha_s/\beta$ at the $t\bar{t}$ threshold where the top-quark
velocity $\beta$ vanishes in the $t\bar{t}$ rest frame, giving rise to
a $K$ factor of the form
\begin{equation}
K_{thr} \to 1+\frac{\alpha_s}{\pi}\, \frac{64}{9} \,
\frac{\pi M_{t}}{\sqrt{(\sqrt{s}-M_H)^2-4 M_t^2}}.
\end{equation}
The pole in $\beta$ is regularized by the vanishing three-particle
phase space for $\beta\to 0$, when the integration for the total cross
section is performed.  This effect can be isolated by studying the
corrections to the Higgs energy spectrum. For reduced Higgs energies,
which shift the $t\bar{t}$ pairs away from the threshold in
$t\bar{t}H$ final states, the corrections are gradually reduced.
Restricting the Higgs energy to 90\% of its kinematically allowed
limit, the $K$ factor is reduced from 1.5 to 1.3 for
$\sqrt{s}=500$~GeV and a Higgs mass of $M_H=120$~GeV.  Since the QCD
corrections are dominated by the virtual corrections and soft-gluon
radiation, the Higgs and top quark energy and angular distributions
are hardly changed, leading to a simple rescaling of the Born cross
section by an approximately uniform $K$ factor.
\\[1em]
\indent At high energies the rescattering corrections become less
important, and the positive contributions from gluon emission are
overwhelmed by the negative Higgs-vertex corrections, as nicely
demonstrated for asymptotic energies in Ref.~\cite{da97}.  For the
dominant mechanism of Higgs radiation off the top quarks, the
reduction of the cross section can be estimated semi-quantitatively in
the limit $M^2_H \ll M^2_t \ll s$. Since the radiation of a low-mass
Higgs boson is in general separated by a large space-time distance
from the top production process, the cross section can crudely be
estimated by the product of the probability for producing a top-quark
pair with the probability for the splitting process $t\to t+H$. The
QCD correction for top-pair production in high-energy $e^+e^-$
annihilation \cite{JLZ} is given by the well-known coefficient
$+\alpha_s/\pi$. The QCD correction to the $Htt$ vertex can easily be
obtained in the low Higgs-mass limit by the logarithmic mass
derivative of the top self-energy \cite{el76,el76a,let}, leading to
$-2\alpha_s/\pi$. Adding up the coefficients $+1-2\times 2=-3$, these
crude estimates suggest a $K$ factor
\begin{equation}
K_{as} \approx 1-3\,\frac{\alpha_s}{\pi}
\end{equation}
in the high-energy limit of the Higgs radiation process; {\it in
numeris} the $K$ factor is therefore predicted to be $K \simeq 0.92$
in the TeV range.
\\[1em]
\indent 
Finally, in Fig.~\ref{fig:cs} we present the cross sections
$\sigma(e^+e^- \rightarrow t \bar{t} H+X)$ for three $e^+e^-$ energy
values $\sqrt{s} = 500$~GeV, 1~TeV and 2~TeV, as a function of the
Higgs mass, in Born approximation and including QCD
corrections.\footnote{The $K$ factors are presented in
Fig.~\ref{fig:K} to simplify experimental simulations of the process.}
Since the Higgs radiation off the top quarks is by far the dominant
subprocess, the cross sections depend approximately quadratically on
the $ttH$ Yukawa coupling so that the experimental sensitivity to the
coupling is very high. This is apparent from Table~\ref{table:cs}
where the contributions from top-quark radiation $(\sigma_{tt})$, $Z$
radiation $(\sigma_{ZZ})$, and the interference term $(\sigma_{tZ})$
are shown separately. While the phase-space suppression gradually
disappears for rising energy, the lower part of the intermediate Higgs
mass range is experimentally accessible already at high-luminosity
$e^+e^-$ colliders for $\sqrt{s} = 500$ GeV; for $\sqrt{s} = 1$ TeV
the entire intermediate Higgs mass range can be covered.  For a
typical size $\sigma \sim 1$ fb of the cross section for the process
$e^+e^- \rightarrow t\bar{t}HX$, about $10^3$ events are generated
when an integrated luminosity $\int {\cal L} \sim 10^3~fb^{-1}$ is
reached within two to three years of running
\cite{ac98}. This should provide a sufficiently large sample for
detailed experimental studies of this process \cite{me98}.
\\[2em]
4. It is straightforward to generalize the present analysis
\cite{di98}.  Within the Standard Model, the calculation of the Higgs
energy spectrum and Higgs--top correlations are required
experimentally. In the context of supersymmetric extensions of the
Standard Model a new calculational element is introduced by the pair
production of scalar and pseudoscalar Higgs bosons \cite{dj92}, while
mixing effects in the scalar Higgs sector merely increase the
complexity of the analysis presented in this note.
\\[2em]
{\small {\bf Acknowledgements}\\
  We thank the authors of Ref.~\cite{da98} for mutual cross checks of
  partial results.  M.K.\ and Y.L.\ are grateful to Prof.~A.~Wagner
  for the warm hospitality extended to them at DESY. We thank
  W.~Beenakker, M.~M\"uhlleitner, and T.~Plehn for many helpful
  discussions.}

\begin{table}[p]
\renewcommand{\arraystretch}{1.2}
\tabcolsep 4pt
\begin{tabular}{|c|c|c|c|c|c|c|}
\hline
\rule[-4mm]{0cm}{1.1cm} 
$\sqrt{s}$ & $M_H$~[GeV] & 
$\sigma_{tt}$~[fb] & 
$\sigma_{ZZ}$~[fb] &
$\sigma_{tZ}$~[fb] &
$\sigma$~[fb] & 
$\displaystyle 
K=\frac{\sigma_{\mathrm{NLO}}}{\vspace*{-1mm}\sigma_{\mathrm{LO}}}$ 
\\ \hline
     & 100 & 1.56 & $2.97 \times 10^{-3}$ & $1.18 \times 10^{-2}$ & 1.58
     & 1.35 \\
 500 GeV & 140 & $8.85 \times 10^{-2}$ & $8.34 \times 10^{-5}$ 
     & $6.76 \times 10^{-4}$ & $8.93 \times 10^{-2}$ & 2.01 \\ 
     & 180 & -- & -- & -- & -- & --\\ \hline
     & 100 & 2.47 & $9.53 \times 10^{-2}$ & $6.15 \times 10^{-2}$ & 2.62
     & 0.95 \\ 
1 TeV & 140 & 1.64 & $8.79 \times 10^{-2}$ & $4.97 \times 10^{-2}$ & 1.78
     & 0.95 \\ 
     & 180 & 1.13 & $7.88 \times 10^{-2}$ & $3.91 \times 10^{-2}$ & 1.24
     & 0.96 \\ \hline
     & 100 & $7.68 \times 10^{-1}$ & $7.91 \times 10^{-2}$ 
     & $2.72 \times 10^{-2}$ & $8.74 \times 10^{-1}$ & 0.90 \\ 
2 TeV & 140 & $5.82 \times 10^{-1}$ & $7.81 \times 10^{-2}$ 
     & $2.57 \times 10^{-2}$ & $6.86 \times 10^{-1}$ & 0.90 \\ 
     & 180 & $4.60 \times 10^{-1}$ & $7.68 \times 10^{-2}$ 
     & $2.41 \times 10^{-2}$ & $5.61 \times 10^{-1}$ & 0.89 \\ \hline
\end{tabular}
\caption[ ]{\it 
 The total cross section for the process $e^+e^-\to t\bar{t} H + X$
 including QCD radiative corrections, split into contributions from 
 Higgs radiation off the top quark $(\sigma_{tt})$, off the $Z$ boson
 $(\sigma_{ZZ})$, and the interference term $(\sigma_{tZ})$, for
 electroweak parameters as specified in the text. Also shown is the
 value of the $K$ factor.}
\label{table:cs}
\end{table}

\newpage


\end{document}